\documentstyle[osa]{revtex}

\begin{document}
\title{Lax Pair Formulation and Multi-soliton Solution of the Integrable Vector$%
\,\, $Nonlinear Schr\"{o}dinger Equation}
\author{Freddy P. Zen$\thanks{%
email: fpzen@bdg.centrin.net.id}^{)1)}\,\,$and \thinspace Hendry I. Elim$%
\thanks{%
email: hendry202@cyberlib.itb.ac.id}^{)1),2)}$}
\author{1)\thinspace Theoretical High Energy Physics Group,}
\author{\thinspace \thinspace \thinspace \thinspace \thinspace \thinspace
Theoretical Physics Lab.,\thinspace Department of Physics,\thinspace}
\author{\thinspace \thinspace \thinspace \thinspace \thinspace \thinspace \thinspace
Institute of Technology Bandung,\thinspace Bandung,\thinspace Indonesia}
\author{2) \thinspace Department of Physics,\thinspace Pattimura University,}
\author{\thinspace \thinspace \thinspace \thinspace \thinspace \thinspace \thinspace
Ambon,\thinspace \thinspace Indonesia}
\maketitle

\begin{abstract}
\thinspace \thinspace \thinspace \thinspace \thinspace \thinspace \thinspace
\thinspace \thinspace \thinspace \thinspace \thinspace \thinspace \thinspace
\thinspace \thinspace \thinspace \thinspace \thinspace \thinspace \thinspace
\thinspace \thinspace \thinspace \thinspace \thinspace \thinspace \thinspace
\thinspace \thinspace \thinspace \thinspace \thinspace \thinspace \thinspace
\thinspace \thinspace \thinspace \thinspace \thinspace \thinspace \thinspace
\thinspace \thinspace \thinspace \thinspace \thinspace \thinspace \thinspace
\thinspace \thinspace \thinspace \thinspace \thinspace \thinspace \thinspace
\thinspace \thinspace \thinspace \thinspace \thinspace \thinspace \thinspace
\thinspace \thinspace \thinspace \thinspace \thinspace \thinspace \thinspace
\thinspace \thinspace \thinspace \thinspace \thinspace \thinspace \thinspace
\thinspace \thinspace \thinspace \thinspace \thinspace \thinspace \thinspace
\thinspace \thinspace \thinspace \thinspace \thinspace \thinspace \thinspace
\thinspace \thinspace \thinspace \thinspace \thinspace \thinspace \thinspace
\thinspace \thinspace \thinspace \thinspace \thinspace \thinspace \thinspace
\thinspace \thinspace \thinspace \thinspace {\bf Abstract}

The integrable vector nonlinear Schr\"{o}dinger (vector {\bf NLS}) equation
is\thinspace investigated by using Zakharov-Shabat ({\bf ZS}) scheme. We get
a Lax pair formulation of the vector {\bf NLS} model. Multi-soliton solution
of the equation is also derived by using inverse scattering method of {\bf ZS%
} scheme. We also find that there is an elastic and inelastic$\,$collision
of the bright and dark multi-solitons of the system. 
\[
\]
PACS number(s): 42.65Sf, 42.65.Tg, 03.40.Kf
\end{abstract}

\section{Introduction}

The integrable coupled nonlinear Schr\"{o}dinger equation of Manakov$^1$
type is widely used in recent developments in the field of optical solitons
in fibers. The use and applications of the equation is to explain how the
solitons waves transmit in optical fiber, what happens when the
interaction\thinspace among optical solitons influences directly the
capacity and quality of communication and so\thinspace on.$^{2-5}$In optical
communications system, the information is coded in binary pulses which
modulate the light carrier wave. We can use bright solitons or dark solitons
in the communications system. The performance of a high-speed fiber
communications system is limited by dispersion, losses, and parasitic
effects induced by electronic repeaters. The use of a single soliton as a
bit of information is interesting and solves the first problem, since the
dispersion broadening effects are compensated for by the nonlinear focusing
effects. The attenuation of solitons due to fiber loss can be compensated
for by amplifying the solitons periodically, in order that they recover
their original width and power. In an all optical communications system,
which does not use electronic repeaters, this amplification can be achieved
by using stimulated Raman scattering.$^6$

The generalization of Manakov type equation to be the integrable vector
nonlinear Schr\"{o}dinger equation (vector {\bf NLS} model) can be written
as follows$^{7-10}$

\begin{equation}
\left( i\frac \partial {\partial x}+\chi \frac{\partial ^2}{\partial t^2}%
+2\mu \sum\limits_{b=1}^m\left| q_b\right| ^2\right)
\,q_c\,=\,0,\,\,\,\,\,\,\,c=1,2,3,...,m\,\,\,\,\,\,,  \eqnum{1.1}
\end{equation}
where $q_c$ are slowly varying envelopes of the $m$ interacting optical
modes, describing a charged field with $m$ colours, the variables $t$ and $x$
are the normalized retarded time and distance along the fiber, $\chi $ and $%
\,\mu \,$are arbitrary real parameters.

The Hamiltonian of the model in eq.$(1.1)$, for $\chi =1$, is$^7$%
\begin{equation}
H=\int \left[ \sum\limits_{c=1}^m\left| \frac{\partial q_c}{\partial t}%
\right| ^2+\mu \left( \sum\limits_{c=1}^m\left| q_c\right| ^2\right)
^2\right] dt,  \eqnum{1.2}
\end{equation}
where integration takes account of the boundary conditions which extend
those for the ordinary {\bf NLS} model.

The set of equations (eq.$(1.1)$, for $\chi =\frac 12$) can also be
described as a propagation of $m$ self-trapped mutually incoherent wave
packets in media with Kerr-like nonlinearity. Related to this case, we can
also interpret that $q_c$ denotes the $c$th component of the beam, $2\mu $
is the coeficient representing the strength of nonlinearity, $t$ is the
transverse coordinate, $x$ is the coordinate along the direction of
propagation, and $\sum\limits_{b=1}^m\left| q_b\right| ^2\,$is the change in
refractive index profile created by all incoherent components of the light
beam. The response time of the nonlinearity is assumed to be long compared
to temporal variations of the mutual phases of all components, so the medium
response to the average light intensity, and this just a simple sum of modal
intensities expressed by the relation $\sum\limits_{b=1}^m\left| q_b\right|
^2.^9$

We \thinspace investigate and derive a Lax pair formulation of the vector
\thinspace {\bf NLS} model using inverse scattering method of {\bf ZS}
scheme. This Lax pair has involved the Lax pair of the integrable singled
and coupled {\bf NLS} equation.$^{5,10}$ After getting the Lax pair, we
solve the equation. We then find multi-soliton solution of the equation. The
solution has involved the solution of the integrable coupled {\bf NLS}
equation of Manakov type appeared in our previous papers in Ref.$5\,$and $10$%
. In this solution, we also get an elastic and inelastic collision of the
bright and dark multi-solitons.

This paper is organized as follows. In section 2, we will perform the {\bf ZS%
} scheme for the Lax pair of the integrable vector nonlinear Schr\"{o}dinger
equation. In section 3, we will solve the bright and dark multi-soliton
solution of the integrable vector {\bf NLS} equation. We also compare our
reduced results with the results in Ref.$8$\thinspace ,$\,9$,$17$ and $%
19\,\, $provided by Seppard and Kivshar, Akhmediev et. al., Radhakrishnan
et. al., and Shchesnovich, respectively. Section 4 is devoted for
discussions and conclusions. In this section, we also state that there must
be a multi-solitons solution in multidimensions ($p+1$) of the integrable
vector {\bf NLS} model.

\section{ZS \thinspace Scheme\thinspace for\thinspace the Lax Pair
Formulation of the Integrable\thinspace \thinspace \thinspace Vector
\thinspace {\bf NLS }Equation}

We start by choosing the following two operators related to Zakharov-Shabat (%
{\bf ZS}) scheme 
\begin{equation}
\Delta _0^{(1)}\,=\,I\left( i\alpha _1\frac \partial {\partial x}\,-\,\frac{%
\partial ^2}{\partial t^2}\right) ,  \eqnum{2.1a}
\end{equation}
$\,$and 
\begin{equation}
\Delta _0^{(2)}\,=\,\left( 
\begin{tabular}{llll}
$\alpha _1$ & 0 & ... & \thinspace \thinspace \thinspace 0 \\ 
0 & $\alpha _2$ & ... & \thinspace \thinspace \thinspace 0 \\ 
... & ... & ... & \thinspace \thinspace ... \\ 
0 & 0 & ... & $\alpha _{m+1}$%
\end{tabular}
\right) \frac \partial {\partial t},  \eqnum{2.1b}
\end{equation}
where $\alpha _1$,$\,\alpha _2$,..., $\alpha _{m+1}\,$ are arbitrary real
values, and $I$ is the $\left( m+1\right) \,$x\thinspace $\left( m+1\right)
\,$unit matrix. We can then define the following\thinspace operators by
using this scheme related to inverse scattering techniques$^{11-13}$%
\begin{equation}
\Delta ^{(1)}\,=\,\,\Delta _0^{(1)}+\,U\left( t,x\right) ,  \eqnum{2.2a}
\end{equation}
and 
\begin{equation}
\Delta ^{(2)}\,=\,\Delta _0^{(2)}+\,V\left( t,x\right) .  \eqnum{2.2b}
\end{equation}
Here operators $\Delta ^{(i)}$,($i\,$=\thinspace 1,\thinspace 2) satisfy the
following equation 
\begin{equation}
\Delta ^{(i)}\left( I\,+\,\Phi _{+}\right) \,=\,\left( I\,+\,\Phi
_{+}\right) \Delta _0^{(i)},  \eqnum{2.3}
\end{equation}
where the Volterra integral operators $\Phi _{\pm }\left( \psi \right) \,$%
\thinspace are defined according to equation 
\begin{equation}
\Phi _{\pm }\left( \psi \right) \,=\,\int\limits_{-\infty }^\infty k_{\pm
}\left( t,z\right) \psi \left( z\right) dz.  \eqnum{2.4}
\end{equation}

We now\thinspace suppose that operators\thinspace $\Phi _F\left( \psi
\right) \,$and\thinspace \thinspace $\Phi _{\pm }\left( \psi \right) \,$ are
related to the following operator\thinspace identity 
\begin{equation}
\left( I\,+\,\Phi _{+}\right) \,\left( I\,+\,\Phi _F\right) \,=\left(
I\,+\,\Phi _{-}\right) ,  \eqnum{2.5}
\end{equation}
where\thinspace the integral operator $\Phi _F\left( \psi \right) \,$%
\thinspace is 
\begin{equation}
\Phi _F\left( \psi \right) \,=\,\int\limits_{-\infty }^\infty F\left(
t,z\right) \psi \left( z\right) dz.  \eqnum{2.6}
\end{equation}
Both $k_{+}\,$and$\,\,F\,\,$in eq.(2.4) and (2.6) are the $\left( m+1\right)
\,$x\thinspace $\left( m+1\right) $ matrices chosen as follows

\begin{equation}
k_{+}=\,\left( 
\begin{tabular}{lllll}
$a\left( t,z;x\right) $ & $q_1\left( t,z;x\right) $ & $q_2\left(
t,z;x\right) $ & ... & $q_m\left( t,z;x\right) $ \\ 
$\pm q_1^{*}\left( t,z;x\right) $ & $d_{11}\left( t,z;x\right) $ & $%
d_{12}\left( t,z;x\right) $ & ... & $d_{1m}\left( t,z;x\right) $ \\ 
$\pm q_2^{*}\left( t,z;x\right) $ & $d_{21}\left( t,z;x\right) $ & $%
d_{22}\left( t,z;x\right) $ & ... & $d_{2m}\left( t,z;x\right) $ \\ 
\thinspace \thinspace \thinspace \thinspace \thinspace \thinspace \thinspace
\thinspace \thinspace \thinspace \thinspace ... & \thinspace \thinspace
\thinspace \thinspace \thinspace \thinspace \thinspace \thinspace \thinspace
... & \thinspace \thinspace \thinspace \thinspace \thinspace \thinspace
\thinspace \thinspace ... & ... & \thinspace \thinspace \thinspace
\thinspace \thinspace \thinspace \thinspace \thinspace \thinspace \thinspace
\thinspace ... \\ 
$\pm q_m^{*}\left( t,z;x\right) $ & $d_{m1}\left( t,z;x\right) $ & $%
d_{m2}\left( t,z;x\right) $ & ... & $d_{mm}\left( t,z;x\right) $%
\end{tabular}
\right) ,  \eqnum{2.7a}
\end{equation}
and 
\begin{equation}
F\,=\,\left( 
\begin{tabular}{lllll}
\thinspace \thinspace \thinspace \thinspace \thinspace \thinspace \thinspace
\thinspace \thinspace \thinspace 0 & $\left( A_1\right) _n$ & $\left(
A_2\right) _n$ & ... & $\left( A_m\right) _n$ \\ 
$\pm \left( A_1^{*}\right) _n$ & \thinspace \thinspace \thinspace \thinspace
0 & \thinspace \thinspace \thinspace \thinspace 0 & ... & \thinspace
\thinspace \thinspace \thinspace \thinspace 0 \\ 
$\pm \left( A_2^{*}\right) _n$ & \thinspace \thinspace \thinspace \thinspace
0 & \thinspace \thinspace \thinspace \thinspace 0 & ... & \thinspace
\thinspace \thinspace \thinspace \thinspace 0 \\ 
\thinspace \thinspace \thinspace \thinspace \thinspace \thinspace \thinspace
\thinspace \thinspace \thinspace ... & \thinspace \thinspace \thinspace
\thinspace ... & \thinspace \thinspace \thinspace \thinspace \thinspace ...
& ... & \thinspace \thinspace \thinspace \thinspace \thinspace ... \\ 
$\pm \left( A_m^{*}\right) _n$ & \thinspace \thinspace \thinspace \thinspace
\thinspace 0 & \thinspace \thinspace \thinspace \thinspace \thinspace 0 & ...
& \thinspace \thinspace \thinspace \thinspace \thinspace \thinspace 0
\end{tabular}
\right) .  \eqnum{2.7b}
\end{equation}
Here $a$, $q_1,q_2,...,q_m$, $\pm q_1^{*},\pm q_2^{*},...,\pm q_m^{*}$, $%
d_{11},d_{12},..,d_{mm}$ , $\left( A_1\right) _n,\left( A_2\right)
_n,...,\left( A_m\right) _n$, and $\left( A_1^{*}\right) _n,\left(
A_2^{*}\right) _n,...,\left( A_m^{*}\right) _n$ (where $n=1,2,3,...,N$) are
parameters which will be calculated in section 3.

In eq.$(2.5)$, we have assumed that $\left( I\,+\,\Phi _{+}\right) \,\,$is
invertible, then 
\begin{equation}
\,\left( I\,+\,\Phi _F\right) \,=\left( I\,+\,\Phi _{+}\right) ^{-1}\left(
I\,+\,\Phi _{-}\right) ,  \eqnum{2.8}
\end{equation}
so that operator $\left( I\,+\,\Phi _F\right) \,$ is factorisable. From eq.$%
(2.5)$, we derive Marchenko matrix equations, 
\begin{equation}
k_{+}\left( t,z;x\right) +\,F\left( t,z;x\right) \,+\,\int\limits_t^\infty
k_{+}\left( t,t^{\prime };x\right) F\left( t^{\prime },z;x\right) dt^{\prime
}\,=\,0,\,\,\,\,\text{\thinspace \thinspace for}\,\,z\,>\,t,\,\, 
\eqnum{2.9a}
\end{equation}
and 
\begin{equation}
k_{-}\left( t,z;x\right) \,=\,F\left( t,z;x\right) \,+\,\int\limits_t^\infty
k_{+}\left( t,t^{\prime };x\right) F\left( t^{\prime },z;x\right) dt^{\prime
}\,,\,\,\,\,\,\,\text{for}\,\,z\,<\,t.\,  \eqnum{2.9b}
\end{equation}
In eq.$(2.9b)$,\thinspace $k_{-}\,\,$is obviously defined in terms of $%
k_{+}\,$and\thinspace $F\,$.\thinspace Both eq.$(2.9a)\,$and $(2.9b)$%
\thinspace require $F$, and $F$ is supplied by the solution of equations : 
\begin{equation}
\left[ \Delta _0^{(1)},\Phi _F\right] =0,  \eqnum{2.10a}
\end{equation}
and 
\begin{equation}
\left[ \Delta _0^{(2)},\Phi _F\right] =0.  \eqnum{2.10b}
\end{equation}
After a little algebraic manipulation, we get 
\begin{equation}
\,i\alpha _1F_x\,+\,F_{zz}\,-\,F_{tt}\,=\,0,  \eqnum{2.11a}
\end{equation}
and 
\begin{equation}
\,\,\,\,\left( 
\begin{tabular}{llll}
$\alpha _1$ & 0 & ... & \thinspace \thinspace \thinspace \thinspace
\thinspace 0 \\ 
0 & $\alpha _2$ & ... & \thinspace \thinspace \thinspace \thinspace
\thinspace 0 \\ 
... & ... & ... & \thinspace \thinspace \thinspace ... \\ 
0 & 0 & ... & $\alpha _{m+1}$%
\end{tabular}
\right) F_t\,+\,F_z\left( 
\begin{tabular}{llll}
$\alpha _1$ & 0 & ... & \thinspace \thinspace \thinspace \thinspace 0 \\ 
0 & $\alpha _2$ & ... & \thinspace \thinspace \thinspace \thinspace 0 \\ 
... & ... & ... & \thinspace \thinspace \thinspace ... \\ 
0 & 0 & ... & $\alpha _{m+1}$%
\end{tabular}
\right) \,=\,0,  \eqnum{2.11b}
\end{equation}
where $F_t\equiv \frac{\partial F}{\partial t}$, $F_z\equiv \frac{\partial F%
}{\partial z}$, etc.

$U\left( t,x\right) \,$and $V\left( t,x\right) \,$can be found by solving eq.%
$(2.3)\,$in which we have substituted eq.($2.7a$) and ($2.4$) (for $k_{+})\,$%
to that equation: 
\begin{equation}
V\left( t,x\right) \,=\left( 
\begin{tabular}{lllll}
\thinspace \thinspace \thinspace \thinspace \thinspace \thinspace \thinspace
\thinspace \thinspace \thinspace \thinspace \thinspace \thinspace \thinspace
\thinspace \thinspace \thinspace \thinspace \thinspace \thinspace \thinspace
0 & $\left( \alpha _1-\alpha _2\right) q_1$ & ... & $\left( \alpha _1-\alpha
_m\right) q_{m-1}$ & $\left( \alpha _1-\alpha _{m+1}\right) q_m$ \\ 
$\pm \left( \alpha _2-\alpha _1\right) q_1^{*}$ & \thinspace \thinspace
\thinspace \thinspace \thinspace \thinspace \thinspace \thinspace \thinspace
\thinspace \thinspace \thinspace \thinspace \thinspace \thinspace 0 & ... & $%
\left( \alpha _2-\alpha _m\right) d_{1\left( m-1\right) }$ & $\left( \alpha
_2-\alpha _{m+1}\right) d_{1m}$ \\ 
$\pm \left( \alpha _3-\alpha _1\right) q_2^{*}$ & $\left( \alpha _3-\alpha
_2\right) d_{21}$ & ... & $\left( \alpha _3-\alpha _m\right) d_{2\left(
m-1\right) }$ & $\left( \alpha _3-\alpha _{m+1}\right) d_{2m}$ \\ 
\thinspace \thinspace \thinspace \thinspace \thinspace \thinspace \thinspace
\thinspace \thinspace \thinspace \thinspace \thinspace \thinspace \thinspace
\thinspace \thinspace \thinspace \thinspace \thinspace \thinspace ... & 
\thinspace \thinspace \thinspace \thinspace \thinspace \thinspace \thinspace
\thinspace \thinspace \thinspace \thinspace \thinspace \thinspace \thinspace
... & ... & \thinspace \thinspace \thinspace \thinspace \thinspace
\thinspace \thinspace \thinspace \thinspace \thinspace \thinspace \thinspace
\thinspace \thinspace \thinspace \thinspace .\thinspace .. & \thinspace
\thinspace \thinspace \thinspace \thinspace \thinspace \thinspace \thinspace
\thinspace \thinspace \thinspace \thinspace \thinspace \thinspace \thinspace
\thinspace ... \\ 
$\pm \left( \alpha _m-\alpha _1\right) q_{m-1}^{*}$ & $\left( \alpha
_m-\alpha _2\right) d_{\left( m-1\right) 1}$ & ... & \thinspace \thinspace
\thinspace \thinspace \thinspace \thinspace \thinspace \thinspace \thinspace
\thinspace \thinspace \thinspace \thinspace \thinspace \thinspace \thinspace
\thinspace 0 & $\left( \alpha _m-\alpha _{m+1}\right) d_{\left( m-1\right)
m} $ \\ 
$\pm \left( \alpha _{m+1}-\alpha _1\right) q_m^{*}$ & $\left( \alpha
_{m+1}-\alpha _2\right) d_{m1}$ & ... & $\left( \alpha _{m+1}-\alpha
_m\right) d_{m\left( m-1\right) }$ & \thinspace \thinspace \thinspace
\thinspace \thinspace \thinspace \thinspace \thinspace \thinspace \thinspace
\thinspace \thinspace \thinspace \thinspace \thinspace \thinspace \thinspace
\thinspace 0
\end{tabular}
\right) ,  \eqnum{2.12a}
\end{equation}
$\,$ and 
\begin{equation}
U\left( t,x\right) \,=\,-2k_{+_t}\,=-2\left( 
\begin{tabular}{lllll}
$a_t$ & $q_{1_t}$ & $q_{2_t}$ & ... & $q_{m_t}$ \\ 
$\pm q_{1_t}^{*}$ & $d_{11_t}$ & $d_{12_t}$ & ... & $d_{1m_t}$ \\ 
$\pm q_{2_t}^{*}$ & $d_{21_t}$ & $d_{22_t}$ & ... & $d_{2m_t}$ \\ 
... & ... & ... & ... & ... \\ 
$\pm q_{m_t}^{*}$ & $d_{m1_t}$ & $d_{m2_t}$ & ... & $d_{mm_t}$%
\end{tabular}
\right) \,.  \eqnum{2.12b}
\end{equation}
Based on the solution of equation $\Delta ^{(2)}\left( I\,+\,\Phi
_{+}\right) \,=\,\left( I\,+\,\Phi _{+}\right) \Delta _0^{(2)}\,$,\thinspace
we get that $k_{+}\left( t,z;x\right) $ must obey the following equation: 
\begin{equation}
\,\,\left( 
\begin{tabular}{llll}
$\alpha _1$ & 0 & ... & \thinspace \thinspace \thinspace \thinspace
\thinspace 0 \\ 
0 & $\alpha _2$ & ... & \thinspace \thinspace \thinspace \thinspace
\thinspace 0 \\ 
... & ... & ... & \thinspace \thinspace \thinspace \thinspace \thinspace ...
\\ 
0 & 0 & ... & $\alpha _{m+1}$%
\end{tabular}
\right) k_{+t}\,+\,k_{+z}\,\,\left( 
\begin{tabular}{llll}
$\alpha _1$ & 0 & ... & \thinspace \thinspace \thinspace \thinspace
\thinspace 0 \\ 
0 & $\alpha _2$ & ... & \thinspace \thinspace \thinspace \thinspace
\thinspace 0 \\ 
... & ... & ... & \thinspace \thinspace \thinspace \thinspace ... \\ 
0 & 0 & ... & $\alpha _{m+1}$%
\end{tabular}
\right) +V\left( t,x\right) \,k_{+}\,=\,0\,,\,\,  \eqnum{2.13}
\end{equation}
and if we evaluate this eq.$(2.13)\,\,$on $z\,=\,t$, we find 
\begin{equation}
a_t\,=\,\mp \frac 1{\alpha _1}\sum\limits_{b=1}^m\left( \alpha _1-\alpha
_{b+1}\right) \mid q_b\mid ^2,\,  \eqnum{2.14a}
\end{equation}
\begin{eqnarray}
d_{11_t} &=&-\frac 1{\alpha _2}\left[ \pm \left( \alpha _2-\alpha _1\right)
\left| q_1\right| ^2+\sum\limits_{b=1}^m\left( \alpha _2-\alpha
_{b+1}\right) d_{1b}d_{b1}\right] ,  \eqnum{2.14b} \\
d_{12_t} &=&-\frac 1{\alpha _2}\left[ \pm \left( \alpha _2-\alpha _1\right)
q_1^{*}q_2+\sum\limits_{b=1}^m\left( \alpha _2-\alpha _{b+1}\right)
d_{1b}d_{b2}\right] ,  \eqnum{2.14c} \\
&&\bullet \,\bullet \,\bullet  \nonumber \\
d_{1m_t} &=&-\frac 1{\alpha _2}\left[ \pm \left( \alpha _2-\alpha _1\right)
q_1^{*}q_m+\sum\limits_{b=1}^m\left( \alpha _2-\alpha _{b+1}\right)
d_{1b}d_{bm}\right] ,  \eqnum{2.14d}
\end{eqnarray}
\begin{eqnarray}
d_{21_t} &=&-\frac 1{\alpha _3}\left[ \pm \left( \alpha _3-\alpha _1\right)
q_2^{*}q_1+\sum\limits_{b=1}^m\left( \alpha _3-\alpha _{b+1}\right)
d_{2b}d_{b1}\right] ,  \eqnum{2.14e} \\
d_{22_t} &=&-\frac 1{\alpha _3}\left[ \pm \left( \alpha _3-\alpha _1\right)
\left| q_2\right| ^2+\sum\limits_{b=1}^m\left( \alpha _3-\alpha
_{b+1}\right) d_{2b}d_{b2}\right] ,  \eqnum{2.14f} \\
&&\bullet \,\bullet \,\bullet  \nonumber \\
d_{2m_t} &=&-\frac 1{\alpha _3}\left[ \pm \left( \alpha _3-\alpha _1\right)
q_2^{*}q_m+\sum\limits_{b=1}^m\left( \alpha _3-\alpha _{b+1}\right)
d_{2b}d_{bm}\right] ,  \eqnum{2.14g}
\end{eqnarray}
\begin{eqnarray}
d_{31_t} &=&-\frac 1{\alpha _4}\left[ \pm \left( \alpha _4-\alpha _1\right)
q_3^{*}q_1+\sum\limits_{b=1}^m\left( \alpha _4-\alpha _{b+1}\right)
d_{3b}d_{b1}\right] ,  \eqnum{2.14h} \\
d_{32_t} &=&-\frac 1{\alpha _4}\left[ \pm \left( \alpha _4-\alpha _1\right)
q_3^{*}q_2+\sum\limits_{b=1}^m\left( \alpha _4-\alpha _{b+1}\right)
d_{3b}d_{b2}\right] ,  \eqnum{2.14i} \\
d_{33_t} &=&-\frac 1{\alpha _4}\left[ \pm \left( \alpha _4-\alpha _1\right)
\left| q_3\right| ^2+\sum\limits_{b=1}^m\left( \alpha _4-\alpha
_{b+1}\right) d_{3b}d_{b3}\right] ,  \eqnum{2.14j} \\
&&\bullet \,\bullet \,\bullet  \nonumber \\
d_{3m_t} &=&-\frac 1{\alpha _4}\left[ \pm \left( \alpha _4-\alpha _1\right)
q_3^{*}q_m+\sum\limits_{b=1}^m\left( \alpha _4-\alpha _{b+1}\right)
d_{3b}d_{bm}\right] ,  \eqnum{2.14k}
\end{eqnarray}
\begin{eqnarray}
&&\bullet \,\bullet \,\bullet  \nonumber \\
d_{m1_t} &=&-\frac 1{\alpha _{m+1}}\left[ \pm \left( \alpha _{m+1}-\alpha
_1\right) q_m^{*}q_1+\sum\limits_{b=1}^m\left( \alpha _{m+1}-\alpha
_{b+1}\right) d_{mb}d_{b1}\right] ,  \eqnum{2.14l} \\
d_{m2_t} &=&-\frac 1{\alpha _{m+1}}\left[ \pm \left( \alpha _{m+1}-\alpha
_1\right) q_m^{*}q_2+\sum\limits_{b=1}^m\left( \alpha _{m+1}-\alpha
_{b+1}\right) d_{mb}d_{b2}\right] ,  \eqnum{2.14m} \\
&&\bullet \,\bullet \,\bullet  \nonumber \\
d_{mm_t} &=&-\frac 1{\alpha _{m+1}}\left[ \pm \left( \alpha _{m+1}-\alpha
_1\right) \left| q_m\right| ^2+\sum\limits_{b=1}^m\left( \alpha
_{m+1}-\alpha _{b+1}\right) d_{mb}d_{bm}\right] .  \eqnum{2.14n}
\end{eqnarray}
The above matrix components can be simplified by choosing $\alpha _2=\alpha
_3=...=\alpha _{m+1}=\delta \,\,$and\thinspace $\alpha _1\neq \delta $,$\,$%
where $\delta \,\,$is an arbitrary real parameter.$\footnote{$^{\#}$The term
which arises for general $\alpha _2,\alpha _3,...,\alpha _{m+1}$ is an
addition term of the vector {\bf NLS }equation.$^{10}$}^{\#}$ We then only
find the following matrix components

\begin{eqnarray}
a_t\, &=&\,\mp \frac{\left( \alpha _1-\delta \right) }{\alpha _1}%
\sum\limits_{b=1}^m\mid q_b\mid ^2,  \eqnum{2.15a} \\
d_{1c_t} &=&\mp \frac{\left( \delta -\alpha _1\right) }\delta \left[
q_1^{*}q_c\right] ,  \eqnum{2.15b} \\
d_{2c_t} &=&\mp \frac{\left( \delta -\alpha _1\right) }\delta \left[
q_2^{*}q_c\right] ,  \eqnum{2.15c} \\
d_{3c_t} &=&\mp \frac{\left( \delta -\alpha _1\right) }\delta \left[
q_3^{*}q_c\right] ,  \eqnum{2.15d} \\
&&\bullet \,\bullet \,\bullet  \nonumber \\
d_{mc_t} &=&\mp \frac{\left( \delta -\alpha _1\right) }\delta \left[
q_m^{*}q_c\right] .  \eqnum{2.15e}
\end{eqnarray}
Plugging the above equations into eq.$(2.12b)$ yields

\begin{equation}
U\left( t,x\right) =-2\left( 
\begin{tabular}{lllll}
$\mp \frac{\left( \alpha _1-\delta \right) }{\alpha _1}\sum\limits_{b=1}^m%
\left| q_b\right| ^2$ & $\,\,\,\,\,\,\,\,\,\,\,\,\,\,\,\,\,\,\,\,q_{1_t}$ & $%
\,\,\,\,\,\,\,\,\,\,\,\,\,\,\,\,\,\,\,q_{2_t}$ & ... & $\,\,\,\,\,\,\,\,\,\,%
\,\,\,\,\,\,\,\,\,\,\,q_{m_t}$ \\ 
$\,\,\,\,\,\,\pm q_{1_t}^{*}$ & $\mp \frac{\left( \delta -\alpha _1\right) }%
\delta \left| q_1\right| ^2$ & $\mp \frac{\left( \delta -\alpha _1\right) }%
\delta \left[ q_1^{*}q_2\right] $ & ... & $\mp \frac{\left( \delta -\alpha
_1\right) }\delta \left[ q_1^{*}q_m\right] $ \\ 
$\,\,\,\,\,\pm q_{2_t}^{*}$ & $\mp \frac{\left( \delta -\alpha _1\right) }%
\delta \left[ q_2^{*}q_1\right] $ & $\mp \frac{\left( \delta -\alpha
_1\right) }\delta \left| q_2\right| ^2$ & ... & $\mp \frac{\left( \delta
-\alpha _1\right) }\delta \left[ q_2^{*}q_m\right] $ \\ 
\thinspace \thinspace \thinspace \thinspace \thinspace \thinspace \thinspace
\thinspace \thinspace \thinspace ... & \thinspace \thinspace \thinspace
\thinspace \thinspace \thinspace \thinspace \thinspace \thinspace \thinspace
\thinspace \thinspace \thinspace ... & \thinspace \thinspace \thinspace
\thinspace \thinspace \thinspace \thinspace \thinspace \thinspace \thinspace
\thinspace \thinspace \thinspace \thinspace ... & ... & \thinspace
\thinspace \thinspace \thinspace \thinspace \thinspace \thinspace \thinspace
\thinspace \thinspace \thinspace \thinspace \thinspace \thinspace \thinspace
... \\ 
$\,\,\,\,\,\pm q_{m_t}^{*}$ & $\mp \frac{\left( \delta -\alpha _1\right) }%
\delta \left[ q_m^{*}q_1\right] $ & $\mp \frac{\left( \delta -\alpha
_1\right) }\delta \left[ q_m^{*}q_2\right] $ & ... & $\mp \frac{\left(
\delta -\alpha _1\right) }\delta \left| q_m\right| ^2$%
\end{tabular}
\right) .  \eqnum{2.16}
\end{equation}
$\,\,\,\,\,\,\,\,\,\,\,\,\,\,\,\,\,\,\,\,\,\,\,\,\,\,\,\,\,\,\,\,\,\,\,\,\,%
\,\,\,\,\,\,\,\,\,\,\,\,\,\,\,\,\,\,\,\,\,\,\,\,\,\,\,\,\,\,\,\,\,\,\,\,\,\,%
\,\,\,\,\,\,\,\,\,\,\,\,\,\,\,\,\,\,\,\,\,\,\,\,\,\,\,\,\,\,\,\,\,\,\,\,\,\,%
\,\,\,\,\,\,$

Now we substitute equations $(2.12a)$ and $(2.16)$ into eq.$(2.2)\,$(for$%
\,\alpha _2=\alpha _3=...=\alpha _{m+1}=\delta \,\,$and\thinspace $\alpha
_1\neq \delta $), we get 
\begin{equation}
\Delta ^{(1)}\,=\,I\left( i\alpha _1\frac \partial {\partial x}\,-\,\frac{%
\partial ^2}{\partial t^2}\right) -2\left( 
\begin{tabular}{lllll}
$\mp \frac{\left( \alpha _1-\delta \right) }{\alpha _1}\sum\limits_{b=1}^m%
\left| q_b\right| ^2$ & $\,\,\,\,\,\,\,\,\,\,\,\,\,\,\,\,\,q_{1_t}$ & $%
\,\,\,\,\,\,\,\,\,\,\,\,\,\,\,\,q_{2_t}$ & ... & $\,\,\,\,\,\,\,\,\,\,\,\,\,%
\,\,\,\,\,q_{m_t}$ \\ 
$\,\,\,\,\,\,\,\,\,\,\,\,\pm q_{1_t}^{*}$ & $\mp \frac{\left( \delta -\alpha
_1\right) }\delta \left| q_1\right| ^2$ & $\mp \frac{\left( \delta -\alpha
_1\right) }\delta \left[ q_1^{*}q_2\right] $ & ... & $\mp \frac{\left(
\delta -\alpha _1\right) }\delta \left[ q_1^{*}q_m\right] $ \\ 
$\,\,\,\,\,\,\,\,\,\,\,\,\pm q_{2_t}^{*}$ & $\mp \frac{\left( \delta -\alpha
_1\right) }\delta \left[ q_2^{*}q_1\right] $ & $\mp \frac{\left( \delta
-\alpha _1\right) }\delta \left| q_2\right| ^2$ & ... & $\mp \frac{\left(
\delta -\alpha _1\right) }\delta \left[ q_2^{*}q_m\right] $ \\ 
\thinspace \thinspace \thinspace \thinspace \thinspace \thinspace \thinspace
\thinspace \thinspace \thinspace \thinspace \thinspace \thinspace \thinspace
\thinspace \thinspace ... & \thinspace \thinspace \thinspace \thinspace
\thinspace \thinspace \thinspace \thinspace \thinspace \thinspace \thinspace
\thinspace \thinspace \thinspace \thinspace \thinspace \thinspace \thinspace
\thinspace ... & \thinspace \thinspace \thinspace \thinspace \thinspace
\thinspace \thinspace \thinspace \thinspace \thinspace \thinspace \thinspace
\thinspace \thinspace \thinspace \thinspace \thinspace \thinspace ... & ...
& \thinspace \thinspace \thinspace \thinspace \thinspace \thinspace
\thinspace \thinspace \thinspace \thinspace \thinspace \thinspace \thinspace
\thinspace \thinspace \thinspace \thinspace \thinspace \thinspace \thinspace
\thinspace ... \\ 
$\,\,\,\,\,\,\,\,\,\,\,\,\pm q_{m_t}^{*}$ & $\mp \frac{\left( \delta -\alpha
_1\right) }\delta \left[ q_m^{*}q_1\right] $ & $\mp \frac{\left( \delta
-\alpha _1\right) }\delta \left[ q_m^{*}q_2\right] $ & ... & $\mp \frac{%
\left( \delta -\alpha _1\right) }\delta \left| q_m\right| ^2$%
\end{tabular}
\right) ,  \eqnum{2.17a}
\end{equation}
and 
\begin{equation}
\Delta ^{(2)}\,=\,\left( 
\begin{tabular}{llll}
$\alpha _1$ & 0 & ... & \thinspace \thinspace \thinspace \thinspace
\thinspace 0 \\ 
0 & $\alpha _2$ & ... & \thinspace \thinspace \thinspace \thinspace 0 \\ 
... & ... & ... & \thinspace \thinspace \thinspace \thinspace ... \\ 
0 & 0 & ... & $\alpha _{m+1}$%
\end{tabular}
\right) \frac \partial {\partial t}\,\,+\,\left( 
\begin{tabular}{lllll}
\thinspace \thinspace \thinspace \thinspace \thinspace \thinspace \thinspace
\thinspace \thinspace \thinspace \thinspace \thinspace \thinspace \thinspace
\thinspace \thinspace \thinspace \thinspace \thinspace \thinspace 0 & $%
\left( \alpha _1-\alpha _2\right) q_1$ & ... & $\left( \alpha _1-\alpha
_m\right) q_{m-1}$ & $\left( \alpha _1-\alpha _{m+1}\right) q_m$ \\ 
$\pm \left( \alpha _2-\alpha _1\right) q_1^{*}$ & \thinspace \thinspace
\thinspace \thinspace \thinspace \thinspace \thinspace \thinspace \thinspace
\thinspace \thinspace \thinspace \thinspace \thinspace 0 & ... & \thinspace
\thinspace \thinspace \thinspace \thinspace \thinspace \thinspace \thinspace
\thinspace \thinspace \thinspace \thinspace \thinspace \thinspace \thinspace
\thinspace \thinspace \thinspace \thinspace \thinspace 0 & \thinspace
\thinspace \thinspace \thinspace \thinspace \thinspace \thinspace \thinspace
\thinspace \thinspace \thinspace \thinspace \thinspace \thinspace \thinspace
\thinspace \thinspace 0 \\ 
$\pm \left( \alpha _3-\alpha _1\right) q_2^{*}$ & \thinspace \thinspace
\thinspace \thinspace \thinspace \thinspace \thinspace \thinspace \thinspace
\thinspace \thinspace \thinspace \thinspace \thinspace 0 & ... & \thinspace
\thinspace \thinspace \thinspace \thinspace \thinspace \thinspace \thinspace
\thinspace \thinspace \thinspace \thinspace \thinspace \thinspace \thinspace
\thinspace \thinspace \thinspace \thinspace \thinspace 0 & \thinspace
\thinspace \thinspace \thinspace \thinspace \thinspace \thinspace \thinspace
\thinspace \thinspace \thinspace \thinspace \thinspace \thinspace \thinspace
\thinspace \thinspace 0 \\ 
$\pm \left( \alpha _4-\alpha _1\right) q_3^{*}$ & \thinspace \thinspace
\thinspace \thinspace \thinspace \thinspace \thinspace \thinspace \thinspace
\thinspace \thinspace \thinspace \thinspace \thinspace 0 & ... & \thinspace
\thinspace \thinspace \thinspace \thinspace \thinspace \thinspace \thinspace
\thinspace \thinspace \thinspace \thinspace \thinspace \thinspace \thinspace
\thinspace \thinspace \thinspace \thinspace \thinspace 0 & \thinspace
\thinspace \thinspace \thinspace \thinspace \thinspace \thinspace \thinspace
\thinspace \thinspace \thinspace \thinspace \thinspace \thinspace \thinspace
\thinspace \thinspace 0 \\ 
\thinspace \thinspace \thinspace \thinspace \thinspace \thinspace \thinspace
\thinspace \thinspace \thinspace \thinspace \thinspace \thinspace \thinspace
\thinspace \thinspace \thinspace \thinspace \thinspace \thinspace \thinspace
... & \thinspace \thinspace \thinspace \thinspace \thinspace \thinspace
\thinspace \thinspace \thinspace \thinspace \thinspace \thinspace \thinspace
\thinspace ... & ... & \thinspace \thinspace \thinspace \thinspace
\thinspace \thinspace \thinspace \thinspace \thinspace \thinspace \thinspace
\thinspace \thinspace \thinspace \thinspace \thinspace \thinspace \thinspace
\thinspace \thinspace ... & \thinspace \thinspace \thinspace \thinspace
\thinspace \thinspace \thinspace \thinspace \thinspace \thinspace \thinspace
\thinspace \thinspace \thinspace \thinspace \thinspace \thinspace ... \\ 
$\pm \left( \alpha _m-\alpha _1\right) q_{m-1}^{*}$ & \thinspace \thinspace
\thinspace \thinspace \thinspace \thinspace \thinspace \thinspace \thinspace
\thinspace \thinspace \thinspace \thinspace \thinspace 0 & ... & \thinspace
\thinspace \thinspace \thinspace \thinspace \thinspace \thinspace \thinspace
\thinspace \thinspace \thinspace \thinspace \thinspace \thinspace \thinspace
\thinspace \thinspace \thinspace \thinspace \thinspace \thinspace 0 & 
\thinspace \thinspace \thinspace \thinspace \thinspace \thinspace \thinspace
\thinspace \thinspace \thinspace \thinspace \thinspace \thinspace \thinspace
\thinspace \thinspace \thinspace \thinspace 0 \\ 
$\pm \left( \alpha _{m+1}-\alpha _1\right) q_m^{*}$ & \thinspace \thinspace
\thinspace \thinspace \thinspace \thinspace \thinspace \thinspace \thinspace
\thinspace \thinspace \thinspace \thinspace \thinspace 0 & ... & \thinspace
\thinspace \thinspace \thinspace \thinspace \thinspace \thinspace \thinspace
\thinspace \thinspace \thinspace \thinspace \thinspace \thinspace \thinspace
\thinspace \thinspace \thinspace \thinspace \thinspace \thinspace 0 & 
\thinspace \thinspace \thinspace \thinspace \thinspace \thinspace \thinspace
\thinspace \thinspace \thinspace \thinspace \thinspace \thinspace \thinspace
\thinspace \thinspace \thinspace \thinspace 0
\end{tabular}
\right) \,.  \eqnum{2.17b}
\end{equation}
Finally, we get a Lax pair of the integrable vector {\bf NLS} equation as
follows

\begin{equation}
i\alpha _1L_x+\left[ L,M\right] =0,  \eqnum{2.18}
\end{equation}
here,

\begin{equation}
L=\Delta ^{\left( 2\right) }=L_0+V\left( t,x\right) ,  \eqnum{2.19a}
\end{equation}
where

\begin{equation}
L_0=\Delta _0^{\left( 2\right) },  \eqnum{2.19b}
\end{equation}
and

\begin{equation}
M=I\left( i\alpha _1\frac \partial {\partial x}\,\right) -\Delta ^{\left(
1\right) }=M_0-U\left( t,x\right) ,  \eqnum{2.19c}
\end{equation}
where

\begin{equation}
M_0=I\frac{\partial ^2}{\partial t^2}.  \eqnum{2.19d}
\end{equation}
$\,\,\,\,\,\,$Since $\Delta ^{(1)}\,$commutes with $\Delta ^{(2)}$\thinspace
, we find the equation which is satisfied by $q_c$ 
\begin{equation}
i\,q_{c_x}+\chi \,q_{c_{tt}}\,+2\mu \sum\limits_{b=1}^m\left| q_b\right|
^2q_c\,=\,0,\,c=1,2,...,m,  \eqnum{2.20a}
\end{equation}
and its complex conjugate

\begin{equation}
-i\,q_{c_x}^{*}+\chi \,q_{c_{tt}}^{*}\,+2\mu \sum\limits_{b=1}^m\left|
q_b\right| ^2q_c^{*}\,=\,0,\,c=1,2,...,m.  \eqnum{2.20b}
\end{equation}
Here parameters $\mu \ $and $\chi \,$are arbitrary real parameters which are
defined as follows

\begin{equation}
\mu \,=\pm \frac{\alpha _1^2-\delta ^2}{\alpha _1^2\delta }\,\,\,, 
\eqnum{2.21a}
\end{equation}
and 
\begin{equation}
\chi =\frac{\left( \alpha _1+\delta \right) }{\alpha _1\left( \alpha
_1-\delta \right) }.  \eqnum{2.21b}
\end{equation}
\thinspace It is obvious that eq.$(2.20a)\,$and $\left( 2.20b\right) $ are a
general form of the integrable vector nonlinear Schr\"{o}dinger equation
(vector {\bf NLS} model) and its complex conjugate.$\,$

\section{The Bright and Dark Multi-Soliton Solution of the Vector {\bf NLS}
Equation}

We consider a general matrix function\thinspace \thinspace $F\,\,$in eq.($%
2.7b$)\thinspace and$\,$substitute\thinspace \thinspace it \thinspace into
eq.$(2.11a)$, we find the following differential equations 
\begin{equation}
i\alpha _1\left( A_c\right) _{n_x}+\left( A_c\right) _{n_{zz}}-\left(
A_c\right) _{n_{tt}}=0,  \eqnum{3.1a}
\end{equation}
and 
\begin{equation}
i\alpha _1\left( A_c^{*}\right) _{n_x}+\left( A_c^{*}\right)
_{n_{zz}}-\left( A_c^{*}\right) _{n_{tt}}=0.  \eqnum{3.1b}
\end{equation}
$\,$The solution of the above$\,\,$equations$\,\,$can be derived by using
separable variable method. We then find 
\begin{equation}
\left( A_c\right) _n\left( t,z;x\right) =\sum\limits_{n=1}^N\left(
A_c\right) _{n_0}e^{-\alpha _1\rho _nz}\left[ e^{\rho _n\left( \delta
t+i\rho _n\left( \alpha _1^2-\delta ^2\right) x\right) }\right] , 
\eqnum{3.2a}
\end{equation}
and

\begin{equation}
\left( A_c^{*}\right) _n\left( t,z;x\right) \,=\,\sum\limits_{n=1}^N\left(
A_c^{*}\right) _{n_0}e^{-\delta \sigma _nz}\left[ e^{\sigma _n\left( \alpha
_1t+i\sigma _n\left( \delta ^2-\alpha _1^2\right) x\right) }\right] , 
\eqnum{3.2b}
\end{equation}
where$\,\sigma _n$,$\,\rho _n\,$,$\left( A_c\right) _{n_0}$and$\,\left(
A_c^{*}\right) _{n_0}$are arbitrary complex parameters.\thinspace \thinspace 
$\,\,\,\,\,\,\,\,\,\,\,\,\,\,\,\,\,\,\,\,\,\,\,\,\,\,\,\,\,\,\,\,\,\,\,\,\,%
\,\,\,\,\,\,\,\,\,\,\,\,\,\,\,\,\,\,\,\,\,\,\,\,\,\,\,\,\,\,\,\,\,\,\,\,\,\,%
\,\,\,\,\,\,\,\,\,\,\,\,\,\,\,\,\,\,\,\,\,\,\,\,\,\,\,\,\,\,\,\,\,\,\,\,\,\,%
\,\,\,\,\,\,\,\,\,\,\,\,\,\,\,\,\,\,\,$

To get the final solution of the integrable vector {\bf NLS} equation, we
have to substitute eq.$(2.7b)$, eq.$(3.2a)\,$and\thinspace \thinspace eq.$%
\left( 3.2b\right) \,$into Marchenko matrix equation (eq.$(2.9a)$). We get
(for $a$, $q_{1\text{,}}\,q_{2\text{,}}\,$..., $q_m$) 
\begin{equation}
a\left( t,z;x\right)
\,=\,-\sum\limits_{n=1}^N\sum\limits_{c=1}^m\int\limits_t^\infty q_c\,\left(
A_c^{*}\right) _{n_0}\exp \left( i\frac{\sigma _n^2}{\alpha _1}\left( \delta
^2-\alpha _1^2\right) x+\sigma _n\left( \alpha _1t^{\prime }-\delta z\right)
\right) dt^{\prime }\,\,,  \eqnum{3.3a}
\end{equation}
and 
\begin{eqnarray}
q_c &=&-\sum\limits_{n=1}^N\left( e^{-\alpha _1\rho _nz}\left( A_c\right)
_{n_0}\right) e^{\delta \rho _nt}e^{i\frac{\rho _n^2}{\alpha _1}\left(
\alpha _1^2-\delta ^2\right) x}  \eqnum{3.3b} \\
&&-\sum\limits_{n=1}^N\left( \left( A_c\right) _{n_0}\right)
\int\limits_t^\infty a\left( t,z;x\right) e^{i\frac{\rho _n^2}{\alpha _1}%
\left( \alpha _1^2-\delta ^2\right) x}e^{-\alpha _1\rho _nz}e^{\delta \rho
_nt^{\prime }}dt^{\prime }.  \nonumber
\end{eqnarray}
The final solution is work on $z=t$. Hence, by substituting eq.$(3.3a)$ to
eq.$(3.3b),\,\,$we find the solution : 
\begin{equation}
q_c=\sum\limits_{n=1}^N\frac{-\left( A_c\right) _{n_0}e^{\rho _n\left(
\delta -\alpha _1\right) t}e^{-i\frac{\rho _n^2}{\alpha _1}\left( \delta
^2-\alpha _1^2\right) x}}{1+\left( \frac{\mu \left(
\sum\limits_{b=1}^m\left| \left( A_b\right) _{n_0}\right| ^2\right) }{\left(
\left( \frac{\alpha _1^2-\delta ^2}{\alpha _1}\right) ^{1/2}\left( \rho
_n-\sigma _n\right) \right) ^2}\right) e^{\rho _n\left( \delta -\alpha
_1\right) t-i\frac{\rho _n^2}{\alpha _1}\left( \delta ^2-\alpha _1^2\right)
x}e^{-\sigma _n\left( \delta -\alpha _1\right) t+i\frac{\sigma _n^2}{\alpha
_1}\left( \delta ^2-\alpha _1^2\right) x}}.  \eqnum{3.4}
\end{equation}
We define

\begin{equation}
\eta _n\,=\,k_n\left( t+ik_n\chi x\right) ,  \eqnum{3.5a}
\end{equation}
and 
\begin{equation}
\eta _n^{*}\,=\,k_n^{*}\left( t-ik_n^{*}\chi x\right) \,\,\,,\,  \eqnum{3.5b}
\end{equation}
where 
\begin{equation}
k_n\,=\,\left( \delta -\alpha _1\right) \rho _n\,\,,\,  \eqnum{3.5c}
\end{equation}
\begin{equation}
k_n^{*}\,=\,-\left( \delta -\alpha _1\right) \sigma _n\,\,,  \eqnum{3.5d}
\end{equation}
and 
\begin{equation}
\tau =\left( \frac{\alpha _1^2-\delta ^2}{\alpha _1}\right) ^{1/2}\left( 
\frac 1{\delta -\alpha _1}\right) .  \eqnum{3.5e}
\end{equation}
Here $\,$\thinspace arbitrary\thinspace complex\thinspace parameter $\,\rho
_n^{*}\,=-\sigma _n.$

Now $q_c$\thinspace \thinspace can be rewritten as 
\begin{equation}
q_c\,=\sum\limits_{n=1}^N\frac{-\left( A_c\right) _{n_0}\,e^{\eta _n}}{%
1\,+\,e^{R_n\,+\,\eta _n\,+\,\eta _n^{*}}},\,\,c=1,2,...,m,  \eqnum{3.6}
\end{equation}
where 
\begin{equation}
e^{R_n}\,=\frac{\mu \left( \sum\limits_{b=1}^m\left| \left( A_b\right)
_{n_0}\right| ^2\right) }{\left( \tau k_n\,+\,\tau k_n^{*}\right) ^2}. 
\eqnum{3.7}
\end{equation}
Based on our solution in eq.$(3.6)$, we can see that our results are the
bright and dark multi-soliton solution since $\mu =\pm \frac{\alpha
_1^2-\delta ^2}{\alpha _1^2\delta }$.\thinspace If we choose parameter $%
\left( \chi \mu \right) $\thinspace $>0$ then we get the bright $N$-solitons
solution. In the case of bright solitons described by the vector {\bf NLS}
equation, the coherent interaction of solitons depends on the relative phase
between them, so that identical solitons with opposite phases repel each
other, whereas in-phase solitons attract each other.$^{14}$On the other
hand, the dark $N$-solitons solution is found when $\left( \chi \mu
\,\right) <\,0$\thinspace . Interaction of the dark solitons is
unconditionally repulsive, in all types of models desribed by the
generalized (vector) {\bf NLS} equation.$^{15}$

The results of $q_c$ show that there are also an elastic and inelastic
collision of the bright and dark multi-soliton. If a nonlinear system
supports propagation of two, or more, waves of different frequencies or
polarization, vector solitons consisting of more than one components can be
formed. Indeed, when two vector solitons are closely separated, they may
form a bound state if the sum of all forces acting between different soliton
components is zero. As an example, let us consider interaction of two vector
solitons consisting of dark and bright components in a defocusing optical
medium. Two dark solitons always repel each other, and as a result they
can't form a bound state.$^{15}$ However, if we introduce out-of-phase
bright components guided by each of the dark solitons, their attractive
interaction creates a proper balance of \thinspace forces, which results in
a stationary two-soliton bound state.

In eq.$(3.6)$, there are several arbitrary complex parameters$\,\left(
A_c\right) _{n_0},\left( A_c^{*}\right) _{n_0}$, $\rho _n\,\,$and$\,\,\sigma
_n$ $\,$which can directly influence the phase of the solitons interaction.
The results of $q_1,q_2,...,q_m$ can also be rewritten in the more
conventional form by introducing $\rho _n\,=l_n+i\lambda _n$ (where $l_n$
and $\lambda _n$ are arbitrary real parameters), 
\begin{equation}
q_c\,=\sum\limits_{n=1}^N\frac{\left( \frac{-\left( \frac{\alpha _1^2-\delta
^2}{\alpha _1}\right) ^{1/2}l_n\left( A_c\right) _{n_0}}{\sqrt{\mu \left(
\sum\limits_{b=1}^m\left| \left( A_b\right) _{n_0}\right| ^2\right) }}%
\right) \exp \left( i\left( \delta -\alpha _1\right) \left[ \lambda
_nt\,+\,\chi \left( \delta -\alpha _1\right) \left( l_n^2-\lambda
_n^2\right) x\right] \right) \,}{\cosh \left[ \left( \delta -\alpha
_1\right) l_n\left( t\,-\,2\chi \left( \delta -\alpha _1\right) \lambda
_nx\,\right) +\,\varphi _n\right] }\,\,,\,\,c=1,2,...,m\,\,,  \eqnum{3.8}
\end{equation}
where $\varphi _n$ is a real multi-soliton phase, 
\begin{equation}
\varphi _n\,=\,\frac 12R_{n,}\,  \eqnum{3.9}
\end{equation}
and the amplitudes of the multi-soliton are

\begin{equation}
Amp^{\left( c\right) }=-\frac{\left( \frac{\alpha _1^2-\delta ^2}{\alpha _1}%
\right) ^{1/2}l_n\left( A_c\right) _{n_0}}{\sqrt{\mu \left(
\sum\limits_{b=1}^m\left| \left( A_b\right) _{n_0}\right| ^2\right) }}%
,c=1,2,...,m.  \eqnum{3.10}
\end{equation}
Parameter $\lambda _n$ contributes to the velocities of the multi-soliton.$%
^{16}$

The result $q_c$ in eq.$(3.8)$ can be reduced to our general multi-soliton
solution of the integrable coupled {\bf NLS} equation of Manakov type
appeared in Ref.$10$\thinspace \thinspace if we put $\chi =\xi ,\,\left(
A_c\right) _{n_0},\,$for $c=1,2\,\,\,$and $\sum\limits_{b=1}^2\left| \left(
A_b\right) _{n_0}\right| ^2$in the equation. On the other hand, we also get
that the bright $N$-solitons solution (for $\left( \chi \mu \right) $ $>0$)
can be reduced to the bright one and two soliton solutions related
to\thinspace that in the works that have been done before by Radhakrishnan,
et.al.\thinspace \thinspace using Hirota method in Ref.$17\,\,$and$%
\,\,18\,\, $\thinspace if we put $\left( A_c\right) _{n_0},\,$for $c=1,2\,$%
\thinspace $\,$and $\sum\limits_{b=1}^2\left| \left( A_b\right)
_{n_0}\right| ^2\,$in eq.$(3.8)$. According to the comparison of the
methods, their \thinspace results\thinspace of\thinspace the bright one
soliton solution\thinspace is equal to our results\thinspace when$\,\alpha
\,e^{\eta _1^{\left( 0\right) }}=\,-\left( A_1\right) _{1_0}\,$,$\,\,\beta
e^{\eta _1^{\left( 0\right) }}=-\left( A_2\right) _{1_0}\,$, $\chi =1$, $%
\tau \simeq 1$, $k_1=\left( \delta -\alpha _1\right) \left( l_1+i\lambda
_1\right) $, $\mu =+\frac{\alpha _1^2-\delta ^2}{\alpha _1^2\delta }\,$and$%
\,\,\left( \mid \alpha \mid ^2+\mid \beta \mid ^2\right) =\left( \left|
\left( A_1\right) _{1_0}\right| ^2+\left| \left( A_2\right) _{1_0}\right|
^2\right) $.\thinspace However,\thinspace their results of the
inelastic\thinspace collision of the bright two soliton\thinspace \thinspace
can be reduced to our\thinspace \thinspace elastic\thinspace \thinspace
collision\thinspace \thinspace of the solution if\thinspace we put $\alpha
_1:\alpha _2=\beta _1:\beta _2$\thinspace in\thinspace their\thinspace
\thinspace result.

Our results can also be reduced to the results in Ref.$19\,\,$provided by
Shchesnovich if we put $\chi =\frac 12,\,2\mu =1,\,x=z,\,t=\tau ,$\thinspace 
$\theta _l=\left( A_c\right) _{n_0},\,l=c=1,2\,,\,n=1,\,$and $\left| \theta
_1\right| ^2+\left| \theta _2\right| ^2=\left| \left( A_1\right)
_{1_0}\right| ^2+\left| \left( A_2\right) _{1_0}\right| ^2=1.\,$In his
results, Shchesnovich had also derived and discussed polarization scattering
by soliton-soliton collisions.

If we compare our multi-solitons solution of the vector {\bf NLS} equation
of Manakov type with the results calculated by Sheppard and Yu. S. Kivshar
in Ref.$8$, we find that their Manakov type results are identical with our
results when $\chi =\frac 12,\,2\mu =-1,\,x=z,\,t=x,$\thinspace and $%
q_c=e_{\pm },\,$for$\,\,c=1,2\,.$\thinspace However, their Hirota method
solutions have been modified and extended in complicated solutions related
to single dark-bright soliton solution and dark-bright multi-soliton
solution.

On the other hand, our multi-solitons solution of the vector {\bf NLS}
equation are identical with the results in Ref.$9$ which is provided by
Akhmediev et. al. when $\chi =\frac 12,\,2\mu =\alpha ,\,x=z,\,t=\tau ,$%
\thinspace $\delta n\left( \psi _i\right) =\sum\limits_{b=1}^{m=N}\left|
q_b\right| ^2,\,$and $q_c=\psi _i,\,$for$\,\,i,c=1,2,...,m\,.$\thinspace
However, Akhmediev et. al. results were derived by using stationary
solutions. Related to the stationary solutions, we can choose the solution
as follows 
\begin{equation}
q_c\left( t,x\right) =\frac 1{\sqrt{2\mu }}G_c\left( t\right) \exp \left( i%
\frac{\theta _c^2}2x\right) ,  \eqnum{3.11}
\end{equation}
then eq.$(1.1)$ can be reduced to 
\begin{equation}
\frac{\partial ^2G_c}{\partial t^2}+2\left[ \sum\limits_{b=1}^mG_b^2\right]
G_c=\theta _c^2G_c,\,\,\,\,\,c=1,2,...,m.  \eqnum{3.12}
\end{equation}
Eq.$(3.12)$ is the same as the related equation in Ref.$9$ when $G_c\left(
t\right) =u_i\left( \tau \right) $ and $\theta _c=\lambda _i.\,$So, we can
generalize that our result is the solution of the bright and dark
multi-solitons collisions of the vector {\bf NLS} model.

\section{Discussions and Conclusions}

We have presented a Lax pair and its bright and dark multi-soliton solution
of the integrable vector {\bf NLS} equation\thinspace using the inverse
scattering Zakharov-Shabat scheme.We can conclude that the solution of the
equation can be solved by using an expanded inverse scattering
Zakharov-Shabat scheme in which we have chosen a certain operator in eq.$%
(2.1)$, and a certain $k_{+}\,\,$(eq.$\left( 2.7a\right) $)$\,$and $F\,$(eq.$%
\left( 2.7b\right) $) in our solution.

Finally, we find that our results correspond to an elastic collision of the
bright and dark vector multi-solitons, as long as 
\begin{equation}
\left[ \left( A_1\right) _{1_0}:\left( A_1\right) _{2_0}:...:\left(
A_1\right) _{N_0}\right] =\left[ \left( A_2\right) _{1_0}:\left( A_2\right)
_{2_0}:...:\left( A_2\right) _{N_0}\right] =...=\left[ \left( A_m\right)
_{1_0}:\left( A_m\right) _{2_0}:...:\left( A_m\right) _{N_0}\right] . 
\eqnum{4.1}
\end{equation}
If in our results, 
\begin{equation}
\left[ \left( A_1\right) _{1_0}:\left( A_1\right) _{2_0}:...:\left(
A_1\right) _{N_0}\right] \neq \left[ \left( A_2\right) _{1_0}:\left(
A_2\right) _{2_0}:...:\left( A_2\right) _{N_0}\right] \neq ...\neq \left[
\left( A_m\right) _{1_0}:\left( A_m\right) _{2_0}:...:\left( A_m\right)
_{N_0}\right] ,  \eqnum{4.2}
\end{equation}
then we get an inelastic collision of the bright and dark vector
multi-solitons. From eq.$(3.8)$, we can conclude that although the collision
between the bright and dark vector multi-solitons, their\thinspace
velocities\thinspace and amplitudes or intensities do not change, their
phases do change. In the eq.($3.8$), we also conclude that the bright and
dark multi-solitons solution of the integrable vector {\bf NLS }equation is
exactly solved by using an extended {\bf ZS} scheme. The applications of our
results can widely contibute to some experiments related to optical fiber in
communications system.

We then propose that there must be an analytical solution for vector
solitons governed by the system of the vector {\bf NLS} model exhibit a
novel type (in Ref.$20$) of solitons collision-the solitons polarization
states switching. Here, the solitons polarization states switching can be
used for the construction of logic elements on bright vector solitons. We
also propose that there is a\thinspace \thinspace vector
multi-solitons\thinspace solution for the other nonlinear equations.
Investigations concerning this problem are now in progress.

\begin{center}
${\bf ACKNOWLEDGEMENTS}$
\end{center}

Both authors would like to thank H.J.Wospakrik for useful discussions. We
also thank N.N. Akhmediev and Yu. S. Kivshar (Optical Sciences Centre, The
Institute for Advanced Studies, The Australian National University ) for
their information and suggestions. We also would like to thank P. Silaban
\thinspace for his encouragements. The work of F.Z. is partially supported
by the Hibah Bersaing VII/1, 1998-1999, DIKTI, Republic of Indonesia.The
work of H.E. is supported by CIDA - EIUDP, Republic of Indonesia.

\end{document}